\documentclass[a4paper,oneside,font=normalsize]{article}
\pdfoutput=1 

\usepackage[textwidth=17cm,textheight=24cm]{geometry}
\usepackage{hyperref}
\usepackage{authblk}
\usepackage[pdftex]{graphicx}
\usepackage[font=small]{caption}   

\usepackage{placeins}

\providecommand{\keywords}[1]{\textbf{\textit{Keywords---}} #1}

\begin{document}
	
	This is an author-created, un-copyedited version of an article accepted for publication in JINST. 
	IOP Publishing Ltd is not responsible for any errors or omissions 
	in this version of the manuscript or any version derived from it. 
	The Version of Record is available online at \url{https://doi.org/10.1088/1748-0221/13/03/C03004}.
	
\title{\textbf{The PANDA Barrel DIRC}}



\author[a]{J.~Schwiening\footnote{Corresponding author, J.Schwiening@gsi.de}} 
\author[a,b]{A.~Ali}
\author[a]{A.~Belias} 
\author[a]{R.~Dzhygadlo} 
\author[a]{A.~Gerhardt} 
\author[a]{K.~G\"{o}tzen} 
\author[a,2]{G.~Kalicy\footnote{present address: The Catholic University of America, Washington, USA}}
\author[a,b]{M.~Krebs}
\author[a]{D.~Lehmann} 
\author[a,b]{F.~Nerling} 
\author[a,3]{M.~Patsyuk\footnote{present address: Massachusetts Institute of Technology, Cambridge, USA}}
\author[a,b]{K.~Peters}
\author[a]{G.~Schepers} 
\author[a,c]{L.~Schmitt}
\author[a]{C.~Schwarz} 
\author[a]{M.~Traxler}
\author[d]{M.~B\"{o}hm}
\author[d]{W.~Eyrich}
\author[d]{A.~Lehmann}
\author[d]{M.~Pfaffinger}
\author[d]{F.~Uhlig}
\author[e]{M.~D\"{u}ren}
\author[e]{E.~Etzelm\"{u}ller}
\author[e]{K.~F\"{o}hl}
\author[e]{A.~Hayrapetyan}
\author[e]{K.~Kreutzfeld}
\author[e]{O.~Merle}
\author[e]{J.~Rieke}
\author[e]{M.~Schmidt}
\author[e]{T.~Wasem}
\author[f]{P.~Achenbach}
\author[f]{M.~Cardinali}
\author[f]{M.~Hoek}
\author[f]{W.~Lauth}
\author[f]{S.~Schlimme}
\author[f]{C.~Sfienti}
\author[f]{M.~Thiel.}

\affil[a]{GSI Helmholtzzentrum f\"ur Schwerionenforschung GmbH, Darmstadt, Germany}
\affil[b]{Goethe University, Frankfurt a.M., Germany}
\affil[c]{FAIR, Facility for Antiproton and Ion Research in Europe, Darmstadt, Germany}     
\affil[d]{Friedrich Alexander-University of Erlangen-Nuremberg, Erlangen, Germany}
\affil[e]{II. Physikalisches Institut, Justus Liebig-University of Giessen, Giessen, Germany}
\affil[f]{Institut f\"{u}r Kernphysik, Johannes Gutenberg-University of Mainz, Mainz, Germany}

\maketitle
\flushbottom

\begin{abstract}

The PANDA experiment at the international accelerator Facility 
for Antiproton and Ion Research in Europe (FAIR) near GSI, Darmstadt, 
Germany will address fundamental questions of hadron physics. 
Excellent Particle Identification (PID) over a large range of 
solid angles and particle momenta will be essential to meet 
the objectives of the rich physics program.
Charged PID for the barrel region of the PANDA target spectrometer 
will be provided by a DIRC (Detection of Internally Reflected 
Cherenkov light) detector.

The Barrel DIRC will cover the polar angle range of 22$^\circ$--140$^\circ$ 
and cleanly separate charged pions from kaons for momenta
between 0.5 GeV/$c$ and 3.5 GeV/$c$ with a separation power of at 
least 3 standard deviations. 
The design is based on the successful BABAR DIRC and the SuperB 
FDIRC R\&D with several important improvements to optimize 
the performance for PANDA, such as a focusing lens system, 
fast timing, a compact fused silica prism as expansion region, 
and lifetime-enhanced Microchannel-Plate PMTs for photon detection.

This article describes the technical design of the PANDA Barrel
DIRC and the result of the design validation using a 
``vertical slice'' prototype in hadronic particle beams 
at the CERN PS.

\end{abstract}

\vspace{10mm}

\keywords{
	Particle identification methods; 
	Cherenkov detectors; 
	Performance of high energy physics detectors.}

\newpage





\normalsize

\section{Introduction}

\begin{figure}[htb]
	\centering
	\includegraphics[width=0.6\textwidth]{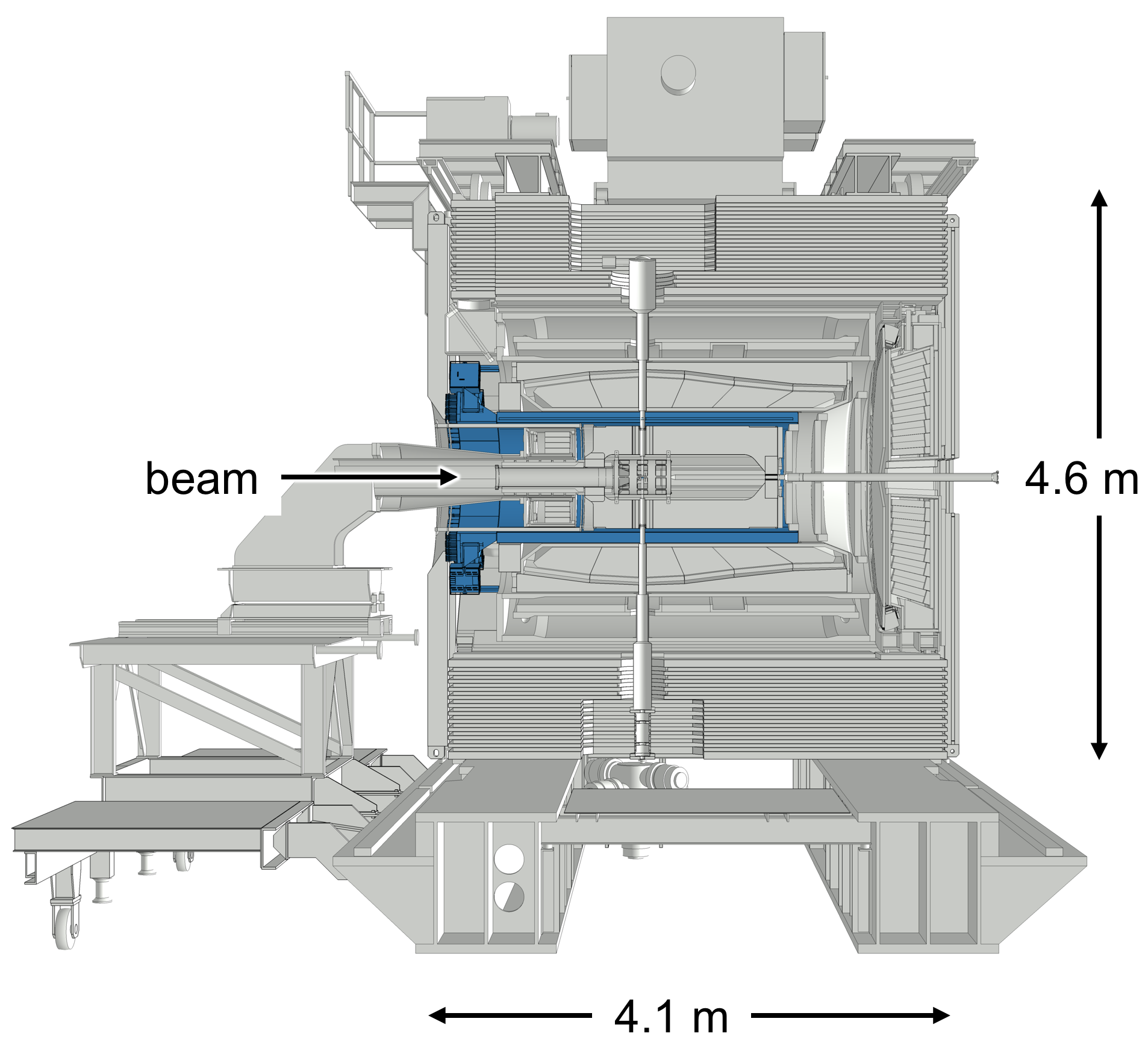} 
	\caption{
		Cross section of the PANDA target spectrometer with the Barrel DIRC marked in blue.
	}
	\label{panda-xsect}
\end{figure}

\begin{figure}[htb]
	\centering
	\includegraphics[width=0.6\textwidth]{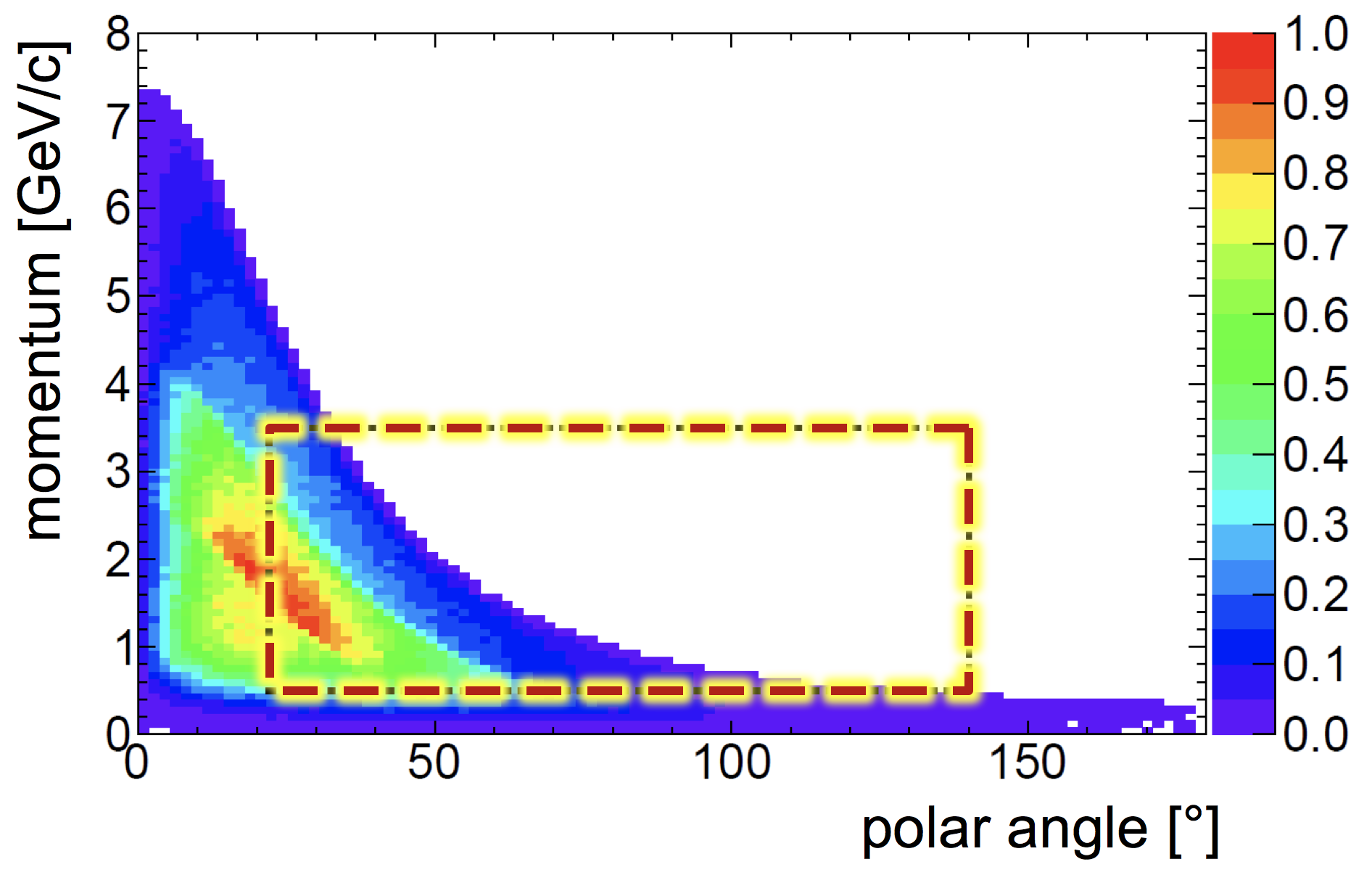}
	\caption{
		Simulated phase space distribution of kaons from
		eight benchmark physics channels at $p_{\bar{p}}$~=~7~GeV/$c$.  
		The Barrel DIRC coverage is marked with the dashed rectangle.
	}
	\label{phasespace}
\end{figure}

The PANDA experiment~\cite{panda1} will be one of the four flagship experiments 
at the new international accelerator complex FAIR (Facility for Antiproton and Ion Research) 
in Darmstadt, Germany.
PANDA will address fundamental questions of hadron
physics and QCD using high-intensity cooled antiproton beams with momenta
between 1.5 and 15~GeV/$c$ and a design luminosity
of up to $2\times10^{32} cm^{-2}s^{-1}$~\cite{panda-physics}.
Excellent Particle Identification (PID) is crucial to the success of the PANDA 
physics program and hadronic PID in the barrel region of the target spectrometer
will be performed by a DIRC (Detection of Internally Reflected
Cherenkov light) counter, see Fig.~\ref{panda-xsect}. 
It is designed to cover the polar angle range from
22$^\circ$ to 140$^\circ$ and provide at least 3~standard deviations (s.d.) $\pi/K$
separation up to 3.5~GeV/$c$, matching the expected upper limit of the 
final state kaon momentum distribution from simulation, see Fig.~\ref{phasespace}.

\section{Barrel DIRC Design}
\label{sec:design}

\begin{figure}[bht]
	\centering
	\includegraphics[width=0.9\textwidth]{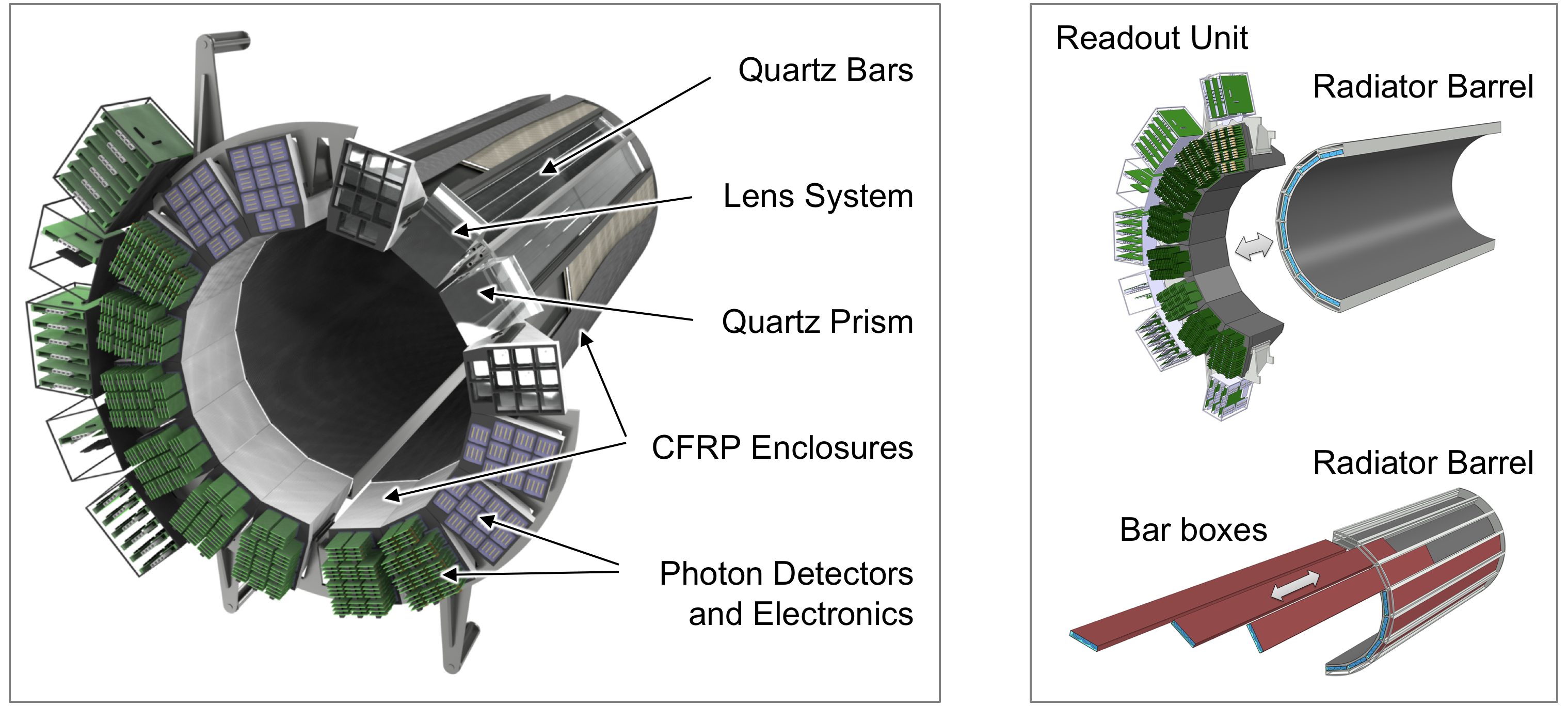}
	\caption{Left: Schematic of the PANDA Barrel DIRC baseline design.
		Right: Mechanical design of the PANDA Barrel DIRC in half-section views,
		highlighting the modular approach.
}
\label{design_3dmec}
\end{figure}

The baseline design of the PANDA Barrel DIRC is shown in Fig.~\ref{design_3dmec}\,(left)
and described in more detail in the Technical Design Report~\cite{TDR-arXiv}.
It is based on the successful BaBar DIRC counter~\cite{adam2005} and key results from 
the R\&D for the SuperB FDIRC~\cite{superb:dirc1} with several
important improvements, such as fast photon timing and a compact imaging region.
The Barrel DIRC consists of sixteen optically isolated sectors, each comprising a 
bar box and a synthetic fused silica (\textit{``quartz''}) prism, surrounding the beam line in a 16-sided 
polygonal barrel with a radius of 476~mm. 
Each bar box contains three synthetic fused silica bars of 17~mm thickness, 53~mm width, 
and 2400~mm length (produced by gluing two 1200~mm-long pieces end-to-end),
placed side-by-side, separated by a small air gap.
A flat mirror is attached to the forward end of each bar to reflect photons towards 
the read-out end, where they are focused by a 3-layer spherical lens on the back 
of a $300$~mm-deep solid prism, made of synthetic fused silica, serving as 
expansion volume (EV).
An array of lifetime-enhanced Microchannel Plate Photomultiplier Tubes 
(MCP-PMTs)~\cite{lehmann2016}, each with 8~$\times$~8 pixels of about 
6.5~$\times$~6.5~mm$^2$ size, is placed at the back surface of the prisms
to detect the photons and measure 
their arrival time on a total of about 11,300 pixels with a precision of $100$~ps 
or better in the magnetic field of approximately 1~T.

The MCP-PMTs are read out by an updated version of the HADES trigger and 
readout board (TRB)~\cite{trb3} in combination with a front-end
amplification and discrimination card (DiRICH)~\cite{dirich} mounted directly on the 
MCP-PMTs.         
This compact FPGA-based system provides measurements of both the photon arrival time and Time-over-Threshold 
(TOT), which is related to the pulse height of the analog signal and can be used to monitor 
the sensor performance and to perform time-walk corrections to improve the 
precision of the photon timing.

The focusing optics has to produce a flat image to match the shape of the back surface 
of the fused silica prism.
This is achieved by a combination of focusing and defocusing elements in a spherical triplet 
lens made from one layer of lanthanum crown glass (NLaK33, refractive index n=1.786 for 
$\lambda$=380~nm) between two layers of synthetic fused silica (n=1.473 for $\lambda$=380~nm).
Such a 3-layer lens works without any air gaps, minimizing the photon loss that would 
otherwise occur at the transition from the lens to the expansion
volume.

The mechanical design divides the Barrel DIRC (Fig.~\ref{design_3dmec}\,(right)) into two main parts: 
The radiator barrel, which contains the radiator bars inside the bar boxes, 
and the readout unit, which includes 
the prism expansion volumes, photon sensors, and Front-End-Electronics (FEE). 
All major mechanical components are expected to be built from aluminum alloy and 
Carbon--Fiber--Reinforced Polymer (CFRP) to minimize the material budget and weight 
and to maximize the stiffness.
The containers for the fused silica bars and prisms are kept under a constant purge 
from boil-off dry nitrogen to maintain a clean and dry environment and avoid possible 
contamination from outgassing of the glue and other materials used in the construction.
The design is modular and allows the installation or removal of each individual 
sealed container holding the optical components.
The readout unit detaches from the radiator barrel for access to the inner 
PANDA detectors during scheduled shutdowns.

This design differs in several key aspects from the version presented
at the DIRC2015 workshop~\cite{schwarz:dirc2015}.
The width of the fused silica bars was increased from 32~mm to 53~mm, reducing the
number of bars per sector from 5 to 3.
The opening angle of the prism was decreased from 38$^\circ$ to 33$^\circ$, which
reduced the total number of MCP-PMTs required from 282 to 176.
These two changes lower the estimated fabrication cost of the 
PANDA Barrel DIRC components by about 20\% without any significant
impact on the performance~\cite{TDR-arXiv}.

\section{Simulation and Reconstruction}
\label{sec:proto}

\begin{figure}[htb]
	\centering
	\includegraphics[width=0.7\textwidth]{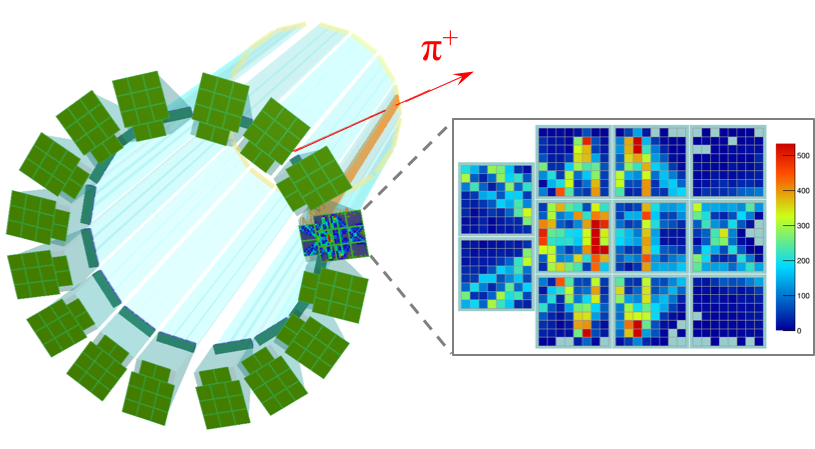}
	\caption{Geant simulation of the PANDA Barrel DIRC baseline design. The path of
		the Cherenkov photons is shown as orange lines.
		The colored histogram shows the accumulated hit pattern from 
		1000~$\pi^+$ at 3.5~GeV/c and $25^\circ$ polar angle.
	}
	\label{design_hp}
\end{figure}

A detailed physical simulation of the PANDA Barrel DIRC was developed in the PANDARoot 
framework~\cite{pandaroot-1,pandaroot-2}, which uses the Virtual Monte Carlo (VMC) 
approach to easily switch between Geant3 and Geant4~\cite{Exc-Geant4} for systematic studies.
The simulation is tuned to the experimentally measured values for the quantum and collection 
efficiency, gain uniformity, and timing resolution of the MCP-PMTs~\cite{fred02}.
Other features include the coefficient of total internal reflection of DIRC radiator bars as a 
function of photon energy~\cite{babar:dirc1}, the bulk transmission of bars, glue, and 
lenses, the wavelength-dependent refractive indices of fused silica, NLaK33, and the 
photocathode, as well as the reflectivity of the forward mirrors.
Figure~\ref{design_hp} shows the Geant representation of the PANDA Barrel DIRC
baseline design together with an example of the accumulated hit pattern produced by 
Cherenkov photons from 1000 $\pi^{+}$ tracks (red line).

Two reconstruction approaches have been developed to evaluate
the detector performance~\cite{RICH14_sim}.
The \emph{geometrical reconstruction}, which is based on the approach used by
the BaBar DIRC~\cite{adam2005}, performs PID by reconstructing
the value of the Cherenkov angle for each detected photon and using it in a 
track-by-track maximum likelihood test.
Furthermore, this algorithm can be used to determine the Cherenkov angle per particle 
and the photon yield, as well as the single photon Cherenkov angle resolution.
While the geometrical method relies mostly on the position of the detected photons
in the reconstruction, the \emph{time-based imaging} approach, based on the method
used by the Belle II TOP~\cite{staric}, utilizes both the position
and time information with comparable precision, and directly performs the 
maximum likelihood test based on the photon arrival time per pixel.

\begin{figure}[htb]
	\centering
	{\includegraphics[width=0.5\textwidth]{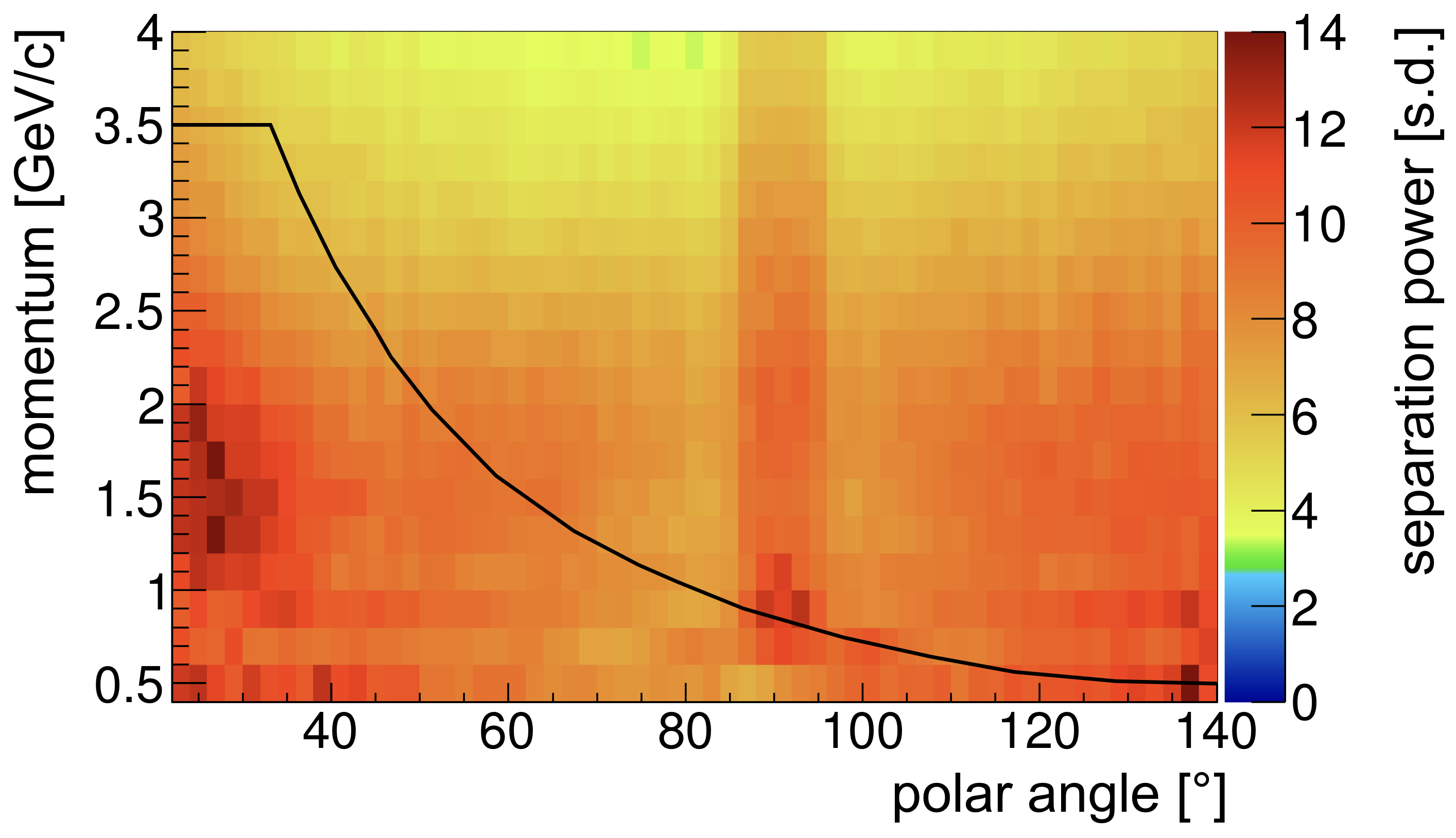}}
	\caption{$\pi$/$K$ separation power as a function of particle momentum and
		polar angle in Geant simulation, determined by the time-based imaging 
		method for the baseline design. 
		The area below the black line corresponds to the final-state phase space
		for charged kaons from various benchmark channels.
	}
	\label{design_simpik}
\end{figure}

The results of the time-based imaging reconstruction of the Geant4 simulation
are shown in Fig.~\ref{design_simpik}.
The $\pi/K$ separation power is shown as a function of the  particle momentum 
and polar angle for a timing precision per photon of 100~ps.
With a separation power of 4--16 s.d., the design exceeds the PANDA PID
requirement for the entire charged kaon phase space,
indicated by the area below the black line.

\section{Prototype Performance in Particle Beams}

A series of increasingly complex PANDA Barrel DIRC system prototypes 
were tested in particle beams at GSI and CERN from 2011--2017 to determine 
the PID performance of various designs and to validate the simulation results. 

\begin{figure}[htb]
	\centering
	\includegraphics[width=0.95\textwidth]{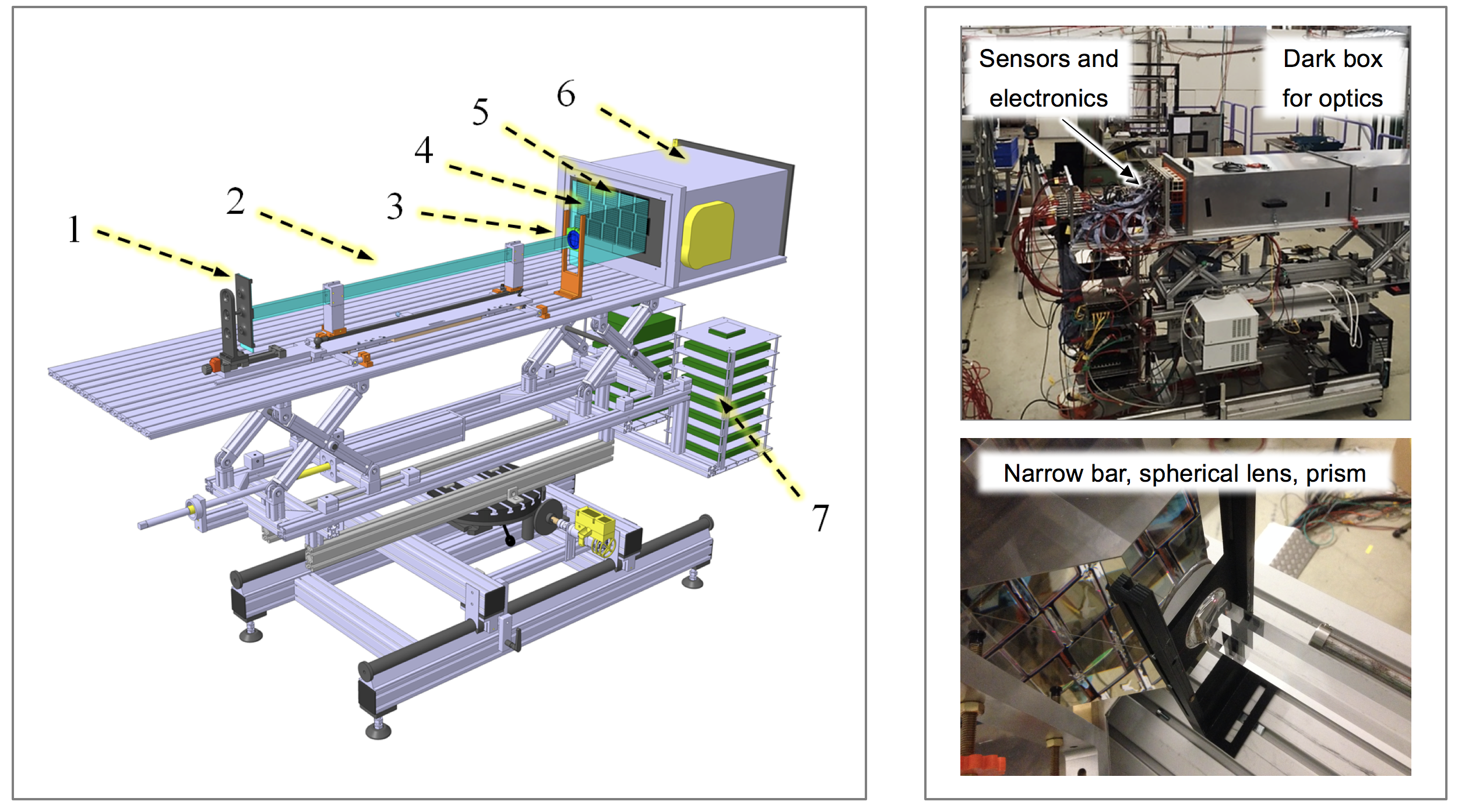}
	\caption{Left: Schematic of the prototype used at CERN in 2015, with 
		1: flat mirror, 2: quartz bar, 
		3: spherical lens, 4: quartz prism, 
		5: array of 5$\times$3 MCP-PMTs, 6: readout unit, and 7: TRB stack.
		Right: Photograph of the 2015 prototype in the CERN T9 beam line (top), close-up
		of the 3-layer spherical lens between the narrow bar and the prism 
		(bottom).
	}
	\label{proto2015}
\end{figure}

The most relevant system prototype test for the validation of the baseline design
took place in 2015 in the T9 beamline at the CERN PS.
The prototype, shown in Fig.~\ref{proto2015}, 
comprised the essential elements of a ``vertical slice'' Barrel DIRC prototype: 
A narrow fused silica bar (17.1 $\times$ 35.9 $\times$ 1200.0~mm$^3$) 
coupled on one end to a flat mirror, on the other end to a 3-layer spherical focusing lens, 
a fused silica prism as expansion volume (with a depth of 300~mm and a 
top angle of 45$^\circ{}$), an array of 3$\times$5 MCP-PMTs, and 1500 readout 
electronics channels using the TRBs in combination with FPGA-based amplification 
and discrimination cards (PADIWA)~\cite{{cardinali:padiwa}}, mounted directly on the MCP-PMTs.

A wide range of data measurements were taken using the mixed hadron beam at different polar angles
and momenta. 
The time difference measured by two time-of-flight stations was used to tag an
event as pion or as proton.
The experimental data showed good agreement of the Cherenkov hit patterns 
with simulation, both in the pixel space and in the photon hit time space.

\begin{figure}[htb]
	\centering
	{\includegraphics[width=0.95\textwidth]{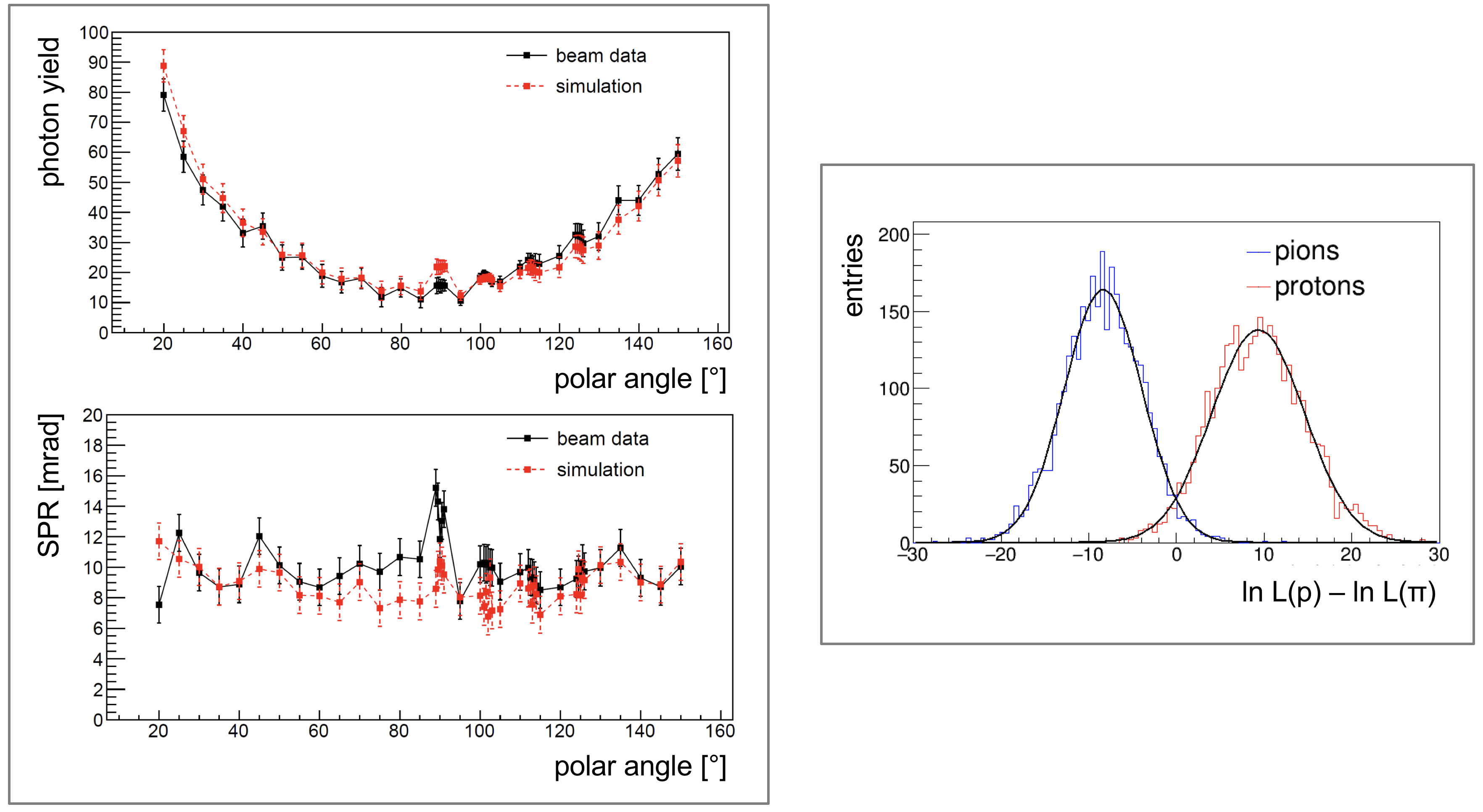}}
	\caption{Performance of the PANDA Barrel DIRC prototype with a narrow bar 
		and a 3-layer spherical lens at the CERN PS in 2015.
		Left: Photon yield (top) and single photon Cherenkov angle resolution (SPR) (bottom) 
		as a function of the track polar angle for tagged protons with a momentum of 7~GeV/$c$.
		The error bars correspond to the RMS of the distribution in each bin. 
		Right: Proton-pion log-likelihood difference distributions for proton-tagged (red) and 
        pion-tagged (blue) beam events as result of the time-based imaging reconstruction
        for a beam with 7~GeV/$c$ momentum and $25^{\circ}$ polar angle.  
	    The $\pi/p$ separation power from the Gaussian fits is 3.6 standard deviations.		
	}
	\label{design_figmerit}
\end{figure}

Examples of the performance of the PANDA Barrel DIRC prototype with a narrow bar 
and a 3-layer spherical lens at the CERN PS in 2015 are shown in Fig.~\ref{design_figmerit}.
The number of Cherenkov photons from the beam data (black) ranges from 12 to 80 
for tagged protons at  7~GeV/$c$ momentum
and is generally in good agreement with simulations (red). 
The single photon Cherenkov angle resolution (SPR) for the same data set 
varies between 9--14~mrad and the beam data and simulation are consistent within the RMS of 
the distributions.
The agreement is noticeably worse for both the photon yield and the SPR at around 
$90^{\circ}$ polar angle,  primarily due to older and lower-gain MCP-PMTs affecting 
this particular angular range.

An example of the log-likelihood difference distributions obtained using time-based 
imaging reconstruction for a polar angle of $25^{\circ}$ 
and a momentum of 7~GeV/$c$ is shown in Fig.~\ref{design_figmerit}\,(right).
Simulation can be used to extrapolate the observed  $\pi/p$  performance of 3.6$\pm$0.1 s.d.
to the expected $\pi/K$ separation power of the fully equipped PANDA Barrel DIRC.
Provided that the expected technical characteristics of the MCP-PMTs, lenses, 
and readout electronics, in particular the photon detection efficiency and 
timing precision, are achieved, the result 
extrapolates to a $\pi/K$ separation power of about 7.7~s.d. at $25^{\circ}$ and 3.5~GeV/$c$ in PANDA.

\section{Conclusion and Outlook}

The technical design of the PANDA Barrel DIRC was completed in 2017 using detailed 
Geant simulations and complex prototypes in particle beams at GSI and CERN.
The key elements are long and narrow bars made from synthetic fused silica as radiators,
spherical focusing lenses, compact fused silica prisms as expansion volumes, 
lifetime-enhanced MCP-PMTs as sensors, and a modular mechanical design.
The performance of this design was validated with hadronic
particle beams and found to meet or exceed the PANDA PID requirements 
for the entire kaon phase space.

The PANDA Barrel DIRC project has now progressed to the construction
phase. 
The schedule foresees the production of the major components,
in particular the fused silica bars and the MCP-PMTs,
to start in 2018 and to be completed in 2021.
After the assembly of the bar boxes and prism boxes, the installation
of the Barrel DIRC into PANDA is expected to take place in 2023,
followed by commissioning with cosmic rays and 	proton beams starting 
in late 2023, leading up to the first PANDA physics run with antiprotons 
in late 2024. 

\section*{Acknowledgments}

This work was supported by HGS-HIRe, HIC for FAIR, and by the
EIC detector R\&D (eRD14) fund,  managed by Brookhaven National Lab.
We thank GSI and CERN staff for the opportunity to use the
beam facilities and for their on-site support.


\end{document}